\documentclass[letterpaper, 10pt, conference]{ieeeconf}
\usepackage{cite}
\usepackage{amsmath,amssymb,amsfonts}
\usepackage{graphicx}
\usepackage{booktabs}
\usepackage{xcolor}
\usepackage{hyperref}
\hypersetup{hidelinks}
\usepackage{amssymb}

\usepackage{amsthm}
\newtheorem{theorem}{Theorem}

\newtheorem{remark}{Remark}

\newcommand{\longthmtitle}[1]{\mbox{}\emph{(#1):}}
\newcommand\xqed[1]{\leavevmode\unskip\penalty9999 \hbox{}\nobreak\hfill\quad\hbox{#1}}
\newcommand\demo{\xqed{$\bullet$}}

\newcommand{\real}{\mathbb{R}}
\newcommand{\Cc}{\mathcal{C}}
\newcommand{\Bc}{\mathcal{B}}
\newcommand{\Kc}{\mathcal{K}}
\newcommand{\bx}{\mathbf{x}}

\newcommand{\bc}{\mathbf{c}}
\newcommand{\bb}{\mathbf{b}}

\newcommand{\bv}{\mathbf{v}}
\newcommand{\bxi}{\boldsymbol{\xi}}
\newcommand{\bA}{\mathbf{A}}
\newcommand{\bB}{\mathbf{B}}
\newcommand{\bK}{\mathbf{K}}
\newcommand{\bI}{\mathbf{I}}
\newcommand{\bS}{\mathbf{S}}

\newcommand{\bp}{\mathbf{p}}
\newcommand{\tp}{\tilde{\bp}}
\newcommand{\xis}{\bxi^{\ast}}
\newcommand{\Az}{\bA_{\zeta}}
\newcommand{\tAk}{\tilde{\bA}_{k\alpha}}
\newcommand{\tbk}{\tilde{\bb}_{\alpha}}
\newcommand{\bzero}{\mathbf{0}}
\newcommand{\tA}{\tilde{\bA}}
\newcommand{\tb}{\tilde{\bb}}

\newcommand{\Ws}{W^{\mathrm{s}}}
\newcommand{\Wu}{W^{\mathrm{u}}}
\newcommand{\gu}{\gamma^{\mathrm{u}}}
\newcommand{\ahom}{\alpha_{\mathrm{hom}}}
\newcommand{\tr}{\operatorname{tr}}
\newcommand{\spec}{\operatorname{spec}}
\newcommand{\norm}[1]{\left\lVert#1\right\rVert}
\newcommand{\setdef}[2]{\{#1 : #2\}}
\DeclareMathOperator*{\argmin}{arg\,min}

\IEEEoverridecommandlockouts
\begin{document}

% Working title. Alternatives kept for the discussion:
%  (b) Safety Filters for Planar Linear Systems: Limit Cycles, Homoclinic Loops, and Regions of Attraction
%  (c) Beyond Global Stability: Regions of Attraction of CBF-Based Safety Filters for Planar Linear Systems
\title{\LARGE \bf Regions of Attraction of CBF-Based Safety Filters \\ for Stable Planar Linear Systems}

\author{Adam K. Kiss, Pol Mestres, Andras Bartfai, Shima Sadat Mousavi, and Aaron D. Ames%
\thanks{\scriptsize Adam K. Kiss, Pol Mestres, Shima Sadat Mousavi, and Aaron D. Ames are with the Department of Mechanical and Civil Engineering, California Institute of Technology, Pasadena, CA 91125, USA.}%
\thanks{\scriptsize Adam K. Kiss and Andras Bartfai are with the Department of Applied Mechanics, Faculty of Mechanical Engineering, Budapest University of Technology and Economics, Budapest, Hungary, {\tt kiss\_a@mm.bme.hu}. Adam K. Kiss was supported by NKFIH (NKKP STARTING 153210) and the Janos Bolyai Research Scholarship of the Hungarian Academy of Sciences. Andras Bartfai is part of the MTA-BME Lendület Machine Tool Vibration Research Group.}%
}

\maketitle
\thispagestyle{empty}

\begin{abstract}
In this paper we study the dynamical properties induced by safety filters based on control barrier functions (CBFs). These are obtained by minimally modifying a nominal controller with the goal of rendering it safe.
We consider the simple setting of a planar linear plant with a stable linear nominal controller and an affine constraint. 
We show that even in this simple setting, the filtered system can oscillate or diverge while remaining safe.
% We provide a full characterization of the set of parameters that lead to a globally (exponentially or asymptotically) stable origin and the set of parameters that do not lead to such dynamical behavior. In the latter case we show that the filter creates a saddle on the constraint boundary, and for underdamped nominal dynamics the boundary of the region of attraction is an unstable periodic orbit, a homoclinic loop, or the saddle's stable manifold, depending on whether the CBF gain is below, at, or above a homoclinic transition value.
We fully characterize the parameters for which the origin is globally (exponentially or asymptotically) stable. For all other parameters, the filter creates a saddle on the constraint boundary, and we characterize the region of attraction of the origin. For underdamped nominal dynamics, its boundary is an unstable periodic orbit, a homoclinic loop, or the saddle's stable manifold, depending on whether the CBF gain is below, at, or above a homoclinic transition value.
A vehicle sideslip example and two three-dimensional examples illustrate the results and what changes beyond planar dynamics.
% A safety filter keeps the state of a controlled system inside a prescribed safe set by substituting the nominal control input with a safe control law when necessary. The simplest planar stable setting is studied, in which this new controller can significantly alter the dynamical behavior. A linear dynamical plant with a linear nominal controller and an affine constraint is considered. Even in this simple scenario the filtered loop can oscillate or diverge while remaining safe, although the plant and the nominal controller are stable. The orientation of the affine constraint boundary determines whether convergence to the origin is global or depends on the initial condition. We establish conditions for global exponential and global asymptotic stability. When the filter creates an additional saddle equilibrium on the constraint boundary, we characterize the region of attraction. For underdamped nominal dynamics, its boundary is an unstable periodic orbit, a homoclinic loop, or the saddle’s stable manifold, depending on how the filter gain compares with a homoclinic transition value. Phase portraits and bifurcation diagrams illustrate these transitions, and a three-state example shows what can change beyond the planar problem.
\end{abstract}

% ================================================================== I
\section{Introduction}

%Many controlled systems must respect hard constraints on their state: an aircraft must not exceed a sideslip or load-factor limit, a vehicle must keep its distance, a robot arm must not reach into a forbidden region. Safety filters are used to minimally modify a nominal desired control law while enforcing satisfaction of the safety constraints. % The task of a safety filter is to ensure that the state never leaves the set on which these limits hold.  Control barrier functions (CBFs)~\cite{AmesTAC2017,AmesECC2019} provide a systematic way to identify control inputs that keep the system within the safe set. % This CBF function is positive inside the allowed safe set, while its value reaches zero on the boundary, together with a condition on the control input that prevents the function from reaching zero.  Although, many inputs may satisfy this condition, the most common and simple way to choose this safety filter is to apply a quadratic program to select the one closest to the nominal command~\cite{AmesTAC2017}. When the nominal input already satisfies the condition, the filter leaves it unchanged.
Many controlled systems must respect hard constraints on their state: an aircraft must not exceed a sideslip or load-factor limit, a vehicle must maintain a safe distance, and a robot arm must avoid forbidden regions. Safety filters provide a systematic way to enforce such constraints by modifying a nominal control input only when necessary. Control barrier functions (CBFs)~\cite{AmesTAC2017,AmesECC2019} characterize control inputs that render a prescribed safe set forward invariant. Since many inputs may satisfy the resulting CBF condition, a common approach is to use a quadratic program (QP) to select the admissible input closest to the nominal command~\cite{AmesTAC2017}. When the nominal input already satisfies the CBF condition, the safety filter leaves it unchanged.

The resulting safety-filtered closed loop is piecewise defined, switching between the nominal controller and the corrective control induced by the filter. Consequently, its global dynamics can differ substantially from those of the nominal closed loop. In~\cite{ReisAguiarTabuada2021,TanDimarogonas2024}, it was shown that CBF-based safety filters can introduce undesired equilibria on the boundary of the safe set, while~\cite{MestresJNLS2026} characterized undesired equilibria together with other spurious behaviors, including unbounded trajectories and limit cycles, for general control-affine systems.

Given the difficulty of characterizing such global phenomena for general nonlinear dynamics and constraints, several works have focused on more structured settings. For planar linear systems with an ellipsoidal safe set,~\cite{ChenMestresCDC2024} 
identified conditions under which the safety boundary has a single saddle equilibrium or multiple equilibria containing one which is asymptotically stable.
%% For linear systems with affine constraints,~\cite{MestresMousaviAmes2026} characterized undesired equilibria and gave sufficient stability conditions, while~\cite{MarcheseECC2025} showed that instability can arise when such conditions fail. Extensions to multiple affine constraints appear in~\cite{MousaviMestresAmesMulti2026}. Frequency-domain and minimum-phase stability conditions were developed in~\cite{LawanCarrascoSu2026,ChoiMinPhase2025}, and~\cite{MestresMousaviOngAmes2026} constructed stabilizing linear feedback laws satisfying the CBF condition without switching.
For linear systems with affine constraints,~\cite{MestresMousaviAmes2026} showed that the closed loop is piecewise affine with two modes, corresponding to the nominal and the filtered input. They characterized the undesired equilibria and gave sufficient conditions for the stability of the desired equilibrium. The four-state example of~\cite{MarcheseECC2025}, in which both modes are stable and the closed loop still oscillates, shows that instability can arise when these conditions fail. Extensions to multiple affine constraints appear in~\cite{MousaviMestresAmesMulti2026}, which characterizes the active-set spectra through invariant zeros, boundary equilibria and tangential unstable directions, and gives geometric and Lyapunov-based conditions for divergence, boundedness, and stability. For linear systems with multiple affine constraints,~\cite{Mousavi2026structure} characterizes QP feasibility, including under input bounds, and derives closed-form safety filters for structured constraint families. For constraints of full relative degree,~\cite{LawanCarrascoSu2026} gave a frequency-domain positive-realness test that ensures global stability, and~\cite{ChoiMinPhase2025} provided minimum-phase conditions for safety-filter stability based on input-output linearization. Building on this framework,~\cite{MestresMousaviOngAmes2026} constructed linear feedback laws without switching that satisfy the CBF condition everywhere.
%These results show that a stable plant with a stable controller can be driven to unexpected behavior (limit cycles, divergence) by the filter, which was added to keep it safe. However, these earlier studies do not characterize the set of trajectories for which such undesired dynamical behaviors arise. In general, as showcased in~\cite{MestresMousaviAmes2026,TanDimarogonas2024,MousaviMestresAmesMulti2026}, even if unbounded trajectories exist, those only exist in a subset of the state space, In this paper, we study this question, along with the size of the region of attraction of the origin (the desired nominal equilibrium), what bounds it, and how it changes with the filter parameters. Given the difficulty of performing such global analysis, in this paper we focus on the case of linear planar systems, for which results such as the Poincar\'e--Bendixson theorem enable such global analysis~\cite{Freire1998,PonceRosVela2022}.

These results establish that a safety filter can create undesired equilibria, limit cycles, or unbounded trajectories even when the nominal closed loop is stable. 
However, a full characterization of the filtered dynamics, as well as the interplay between such dynamics and the safety filter parameters is still missing, even for most simple classes of systems and safety constraints.
Moreover, the existing results ensuring global stability of the closed-loop system do not reveal which initial conditions exhibit them. In particular, unbounded trajectories may coexist with trajectories that still converge to the desired equilibrium~\cite{MestresMousaviAmes2026,TanDimarogonas2024,MousaviMestresAmesMulti2026}. 

This motivates the central question of this paper: \emph{what is the region of attraction of the desired equilibrium, what forms its boundary, when is its safe part bounded, and how do these depend on the safety-filter parameters?}
Addressing this question requires a global characterization of the filtered dynamics. We therefore focus on planar linear systems and affine CBF constraints. The planar nature of the dynamics enables the use of tools from planar dynamical systems, including the Poincar\'e--Bendixson theorem, which make such an analysis possible~\cite{Freire1998,PonceRosVela2022}.

%The contributions are as follows. First, the planar closed-loop is reduced to a normal form which only depends on three parameters. Namely, the damping ratio $\zeta$ of the nominal loop (which we assume to be stable), a dimensionless parameter $k$ of the constraint whose sign decides whether the mode the filter activates is stable, and the filter gain $\alpha$. Second, we prove that for $k>0$ the origin is globally exponentially stable (GES) for every gain, improving on the sufficient condition given in~\cite{MestresMousaviAmes2026}. For $k=0$, we prove global asymptotic (but not exponential) stability of the origin. Third, for $k<0$ we show that the filter creates a saddle on the boundary of the safe set, and provide a full characterization of the region of attraction of the origin:  it is an unstable periodic orbit at small gains, a homoclinic loop at a homoclinic transition gain determined by an explicit scalar equation, and the stable manifold of the saddle above it. For a nominal node, the stable manifold bounds the basin for every gain.  Finally, we present two examples illustrating the greater complexity of three-dimensional safety filters, whose dynamics can exhibit phenomena such as stable limit cycles and chaos.
The contributions are as follows. First, we reduce the planar safety-filtered closed loop to a dimensionless normal form with three parameters: the damping ratio $\zeta$ of the stable nominal dynamics, a normalized invariant-zero parameter $k$, whose sign determines the stability of the active mode, and the filter gain $\alpha$. Second, for $k>0$, we prove global exponential stability of the origin for every $\alpha>0$, strengthening the global stability result in~\cite{MestresMousaviAmes2026}. For $k=0$, we establish global asymptotic, but not exponential, stability. Third, for $k<0$, the filter creates a saddle equilibrium on the constraint boundary, and we characterize the region of attraction of the origin. For underdamped nominal dynamics, its boundary is an unstable periodic orbit at small gains, a homoclinic loop at a unique transition gain given by an explicit scalar equation, and the saddle's stable manifold above this transition. For critically damped and overdamped dynamics, the stable manifold bounds the region of attraction for every positive gain. Finally, we illustrate the results on a vehicle sideslip-protection problem and with three-dimensional examples showing behavior beyond the planar setting.

%The rest of the paper is organized as follows. Section~\ref{sec:model} sets up the closed loop, the structure of its active mode, and its normal form. Section~\ref{sec:results} states and proves the planar results and illustrates their bifurcations. Section~\ref{sec:vehicle} applies the planar results to vehicle sideslip protection, Section~\ref{sec:3d} presents the three-state examples, and Section~\ref{sec:conclusion} concludes.

% ================================================================== II
\section{Background}\label{sec:model}

\subsection{Safety filters}

A state constraint is represented by a continuously differentiable function $h$ and the corresponding safe set
\begin{equation}\label{eq:safeset}
    \Cc := \setdef{\bx\in\real^n}{h(\bx)\geq 0}.
\end{equation}
For a control-affine system $\dot\bx=f(\bx)+g(\bx)u$, any locally Lipschitz controller satisfying the CBF condition\footnote{Where $\kappa$ is an extended class-$\Kc$ function. We use the linear choice $\kappa(s)=\alpha s$ with $\alpha>0$.}
\begin{equation}\label{eq:cbf}
    \dot h(\bx,u)+\alpha h(\bx)\geq0
\end{equation}
on $\Cc$ renders $\Cc$ forward invariant~\cite{AmesTAC2017}.

Given a nominal controller $u_{\rm nom}$, a CBF-based safety filter selects the admissible input closest to the nominal command:
\begin{equation}\label{eq:qpgen}
    u^{\ast}(\bx)=\argmin_{u}\ \tfrac12\norm{u-u_{\rm nom}(\bx)}^2
    \quad\text{s.t.}\quad \eqref{eq:cbf}.
\end{equation}
% In the relative-degree-one setting considered below, this QP has a single affine constraint 
The Karush-Kuhn-Tucker (KKT) conditions yield $u^\ast$ in closed form \cite{Kiss_CBFal}. 
% The multiplier vanishes where the nominal input already satisfies~\eqref{eq:cbf}, and elsewhere~\eqref{eq:cbf} holds with equality~\cite{XuRobust2015}. The filter switches between two control laws, and this non-smoothness makes the closed loop inherently nonlinear.

\subsection{Planar linear plant with an affine constraint}

We now specialize the safety filter~\eqref{eq:qpgen} to a planar linear plant with a scalar input and a linear nominal controller,
\begin{equation}\label{eq:plant}
    \dot{\bx} = \bA\bx + \bB u, \qquad
    u_{\rm nom}(\bx) = -\bK\bx,
    \qquad \bx\in\real^2,\ u\in\real,
\end{equation}
subject to the affine constraint
\begin{equation}\label{eq:barrier}
    h(\bx) = \bc^\top\bx + d .
\end{equation}
Unless stated otherwise, we assume throughout that
\begin{itemize}
    \item the nominal closed-loop matrix
    \begin{equation}\label{eq:A0}
        \bA_0 := \bA-\bB\bK
    \end{equation}
    is Hurwitz;
    \item $d>0$, so the origin lies in the interior of the safe set;
    \item $\bc^\top\bB\neq0$, so that $h$ has relative degree one.
\end{itemize}

Under the nominal controller, the left-hand side of the CBF condition~\eqref{eq:cbf} is
\begin{equation}\label{eq:eta}
    \eta(\bx)
    :=\bc^\top(\bA_0+\alpha\bI)\bx+\alpha d,
\end{equation}
which we call the nominal CBF margin. If $\eta(\bx)\geq0$, the nominal input is feasible and the filter is inactive. If $\eta(\bx)<0$, the KKT conditions of~\eqref{eq:qpgen} make the CBF constraint active, yielding
\[
    u^\ast(\bx)
    =-\bK\bx-\frac{\eta(\bx)}{\bc^\top\bB}.
\]
Substituting this expression into~\eqref{eq:plant} gives the closed-loop dynamics
\begin{equation}\label{eq:cl}
    \begin{gathered}
    \dot{\bx} =
    \begin{cases}
        \bA_0\bx, & \eta(\bx)\geq 0,\\[2pt]
        \tA\bx+\tb, & \eta(\bx)<0,
    \end{cases}
    \\[3pt]
    \begin{aligned}
    \tA &:= \bA_0
    -\frac{\bB\,\bc^\top(\bA_0+\alpha\bI)}
    {\bc^\top\bB},\\
    \tb &:= -\frac{\alpha d}{\bc^\top\bB}\,\bB .
    \end{aligned}
    \end{gathered}
\end{equation}
The switching set
\[
    \Sigma:=\setdef{\bx}{\eta(\bx)=0}
\]
separates the inactive zone $\{\eta\geq0\}$, where $u=u_{\rm nom}$, from the active zone
$\{\eta<0\}$. The two vector fields in~\eqref{eq:cl} coincide on
$\Sigma$, so the closed-loop vector field is continuous and piecewise
affine.

\subsection{Equilibria and their stability}

The origin is an equilibrium of~\eqref{eq:cl}. Indeed,
$\eta(\bzero)=\alpha d>0$, so the filter is inactive in a
neighborhood of the origin, and local exponential stability follows
from $\bA_0$ being Hurwitz.

To characterize possible additional equilibria, consider the active
dynamics. Since the filter enforces
$\dot h=-\alpha h$ in the active zone,
\[
    \bc^\top\tA=-\alpha\bc^\top.
\]
Hence $-\alpha$ is an eigenvalue of $\tA$, while the second eigenvalue
is
\begin{equation}\label{eq:spec}
    \spec\tA=\{-\alpha,z\},\qquad
    z:=\tr\bA-\frac{\bc^\top\bA\bB}{\bc^\top\bB}.
\end{equation}
The eigenvalue $-\alpha$ governs motion normal to the constraint
boundary, whereas $z$ governs the dynamics tangential to it.
Moreover, $z$ is the invariant zero of the SISO system
$(\bA,\bB,\bc^\top)$. This spectral characterization is the planar
single-input specialization of~\cite[Lemma~3]{MestresMousaviAmes2026}
(cf.~\cite[Remark~1]{MestresMousaviAmes2026}).

For $z\neq0$, the active matrix $\tilde{\bA}$ is invertible, and the affine active
dynamics have the unique candidate equilibrium
\begin{equation}\label{eq:ptil}
    \tp:=-\tA^{-1}\tb,\qquad h(\tp)=0.
\end{equation}
The condition $h(\tp)=0$ follows directly from
$\dot h=-\alpha h$ at equilibrium. This candidate is an equilibrium
of the full filtered system only if it lies in the active zone,
i.e., if $\eta(\tp)<0$. Direct substitution gives
\begin{equation}\label{eq:etaptil}
    \eta(\tp)=-\frac{d\,\det\bA_0}{z}.
\end{equation}
Since $d>0$ and $\det\bA_0>0$, this occurs if and only if $z>0$.
Thus, for $z<0$ the origin is the only equilibrium, whereas for
$z>0$ the filtered system has an additional equilibrium $\tp$ on
the safety boundary. This recovers the planar case of
\cite[Prop.~5]{MestresMousaviAmes2026}. The singular case $z=0$ is
treated in Section~\ref{sec:ges}.

The same sign distinction determines the stability of the active
mode. If $z<0$, then $\tA$ is Hurwitz. If $z>0$, $\tp$ is a saddle
with eigenvalues $-\alpha<0$ and $z>0$. Since
$\bc^\top\tA=-\alpha\bc^\top$, an eigenvector associated with $z$ is
orthogonal to $\bc$ and therefore tangent to the constraint boundary
$\{h=0\}$. Thus, the unstable direction of the active mode is
tangential to the safety boundary.

\subsection{Normal form}\label{sec:normal}

The closed loop~\eqref{eq:cl} is set by twelve parameters: the entries of $\bA$, $\bB$, $\bK$ and $\bc$, the offset $d$, and the gain $\alpha$. Next we show how a change of coordinates and time scale reduces them to three.

Let $\omega_{\rm n}:=\sqrt{\det\bA_0}$ and $\zeta:=-\tr\bA_0/(2\omega_{\rm n})$, both positive since $\bA_0$ is Hurwitz, and let
\begin{equation}\label{eq:S}
    \bS := \Bigl[\ \frac{(\bA_0-(\tr\bA_0)\bI)\bB}{\omega_{\rm n}\,\bc^\top\bB}\ \Big|\ \frac{\bB}{\bc^\top\bB}\ \Bigr],
\end{equation}
which is invertible when $(\bA,\bB)$, equivalently $(\bA_0,\bB)$, is controllable~\cite[Thms.~8.1--8.2]{Chen1999}, as we assume from here on. The coordinates $\bxi=(\xi_1,\xi_2)^\intercal$ defined by $\bx=\bS\bxi$ are the controllable canonical form of $(\bA_0,\bB)$, scaled so that the constraint has unit coefficient on $\xi_2$. A direct computation gives $\bS^{-1}\bA_0\bS=\omega_{\rm n}\Az$, $\bS^{-1}\bB=(\bc^\top\bB)(0,1)^\top$ and $\bS^\top\bc=(k,1)^\top$, with $k:=-z/\omega_{\rm n}$ and $\Az$ as below. In these coordinates and in the dimensionless time $\tau=\omega_{\rm n} t$, the plant and the constraint read
\begin{equation}\label{eq:nf}
    \begin{gathered}
    \frac{d\bxi}{d\tau} = \Az\,\bxi + \begin{bmatrix} 0 \\ 1\end{bmatrix} u,
    \quad
    \Az := \begin{bmatrix} 0 & 1 \\ -1 & -2\zeta\end{bmatrix},
    \\[2pt]
    h = k\,\xi_1 + \xi_2 + d .
    \end{gathered}
\end{equation}
with nominal controller $u=0$. 
With a slight abuse of notation, we let $\alpha$ be the dimensionless gain $\alpha/\omega_{\rm n}$. The offset $d$ is a scale, since the solution through $d\,\bxi_0$ at offset $d$ is $d$ times the solution through $\bxi_0$ at offset $1$. Three parameters are left defining the system dynamics: $\zeta$, $k$ and $\alpha$.

The nominal loop in~\eqref{eq:nf} is a second-order oscillator with damping ratio $\zeta$ and normalized poles $\lambda_{1,2}=-\zeta\pm\sqrt{\zeta^2-1}$. 
When $0<\zeta<1$, the system is underdamped and has a complex conjugate pair. When $\zeta = 1$ it is critically damped and has a repeated negative real pole, and when $\zeta > 1$ it is overdamped and has two distinct negative real poles.
The undamped limit ($\zeta=0$) has purely imaginary poles $\pm i$. The constraint boundary is the line $\xi_2=-k\xi_1-d$, so $-k$ defines its slope in the phase plane and $\varphi=\arctan k\in(-\pi/2,\pi/2)$ its angle, and $k$ measures the zero of the constraint in units of the natural frequency. Its sign is the sign of $-z$. The active mode is Hurwitz for $k>0$ and a saddle for $k<0$.

In these coordinates the closed loop~\eqref{eq:cl} reads
\begin{equation}\label{eq:nfcl}
    \begin{gathered}
    \frac{d\bxi}{d\tau} =
    \begin{cases}
        \Az\bxi, & \eta(\bxi)\geq 0,\\[2pt]
        \tAk\bxi+\tbk, & \eta(\bxi)<0,
    \end{cases}
    \\[3pt]
    \tAk = \begin{bmatrix} 0 & 1\\ -k\alpha & -(k{+}\alpha)\end{bmatrix},\quad
    \tbk = \begin{bmatrix} 0\\ -\alpha d\end{bmatrix}.
    \end{gathered}
\end{equation}
with $\spec\tAk=\{-\alpha,-k\}$, while the margin~\eqref{eq:eta} reads $\eta(\bxi)=(k,1)(\Az+\alpha\bI)\bxi+\alpha d$. For $k<0$ the saddle $\tp$ of~\eqref{eq:ptil} is
\begin{equation}\label{eq:xis}
    \xis = \bigl(-d/k,\ 0\bigr)^\top , \qquad h(\xis)=0,\qquad \eta(\xis)=d/k,
\end{equation}
with eigenvectors $\bv_k=(1,-k)^\top$ for $-k$ and $\bv_\alpha=(1,-\alpha)^\top$ for $-\alpha$. The first is parallel to the constraint boundary. Section~\ref{sec:results} works in the normal form. To return to the original problem, states are easily mapped back by $\bx=\bS\bxi$.

% ================================================================== III
\section{Main results}\label{sec:results}

In this section we seek to characterize the region of attraction of the origin as well as the set of initial conditions that lead to unbounded trajectories.
We denote by $\Bc(\bzero)$ the region of attraction of the origin and refer to $\Bc(\bzero)\cap\Cc$ as the \emph{safe recoverable set}. 
We consider the cases $k \geq 0$ and $k < 0$ separately.

% Assuming positive nominal damping, $k>0$ gives global exponential recovery, $k=0$ gives global asymptotic recovery, and $k<0$ makes recovery depending on the initial condition. First, the global result are established, then the basin boundaries are identified depicted in Fig.~\ref{fig:cell}, and finally the zero-damping limit is discussed.

\begin{figure*}[t]
    \centering
    \includegraphics[width=\textwidth]{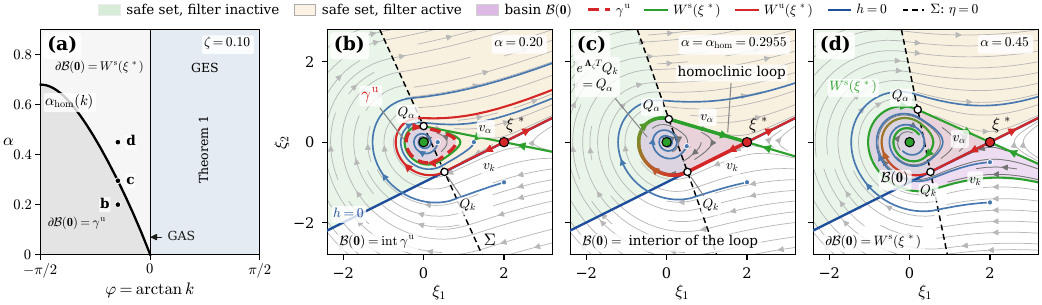}
%    \caption{Basin boundaries for $\zeta=0.1$, $d=1$. (a) Parameter map. The three marked gains at $k=-0.5$ correspond to: (b) an unstable periodic orbit, $\alpha=0.20$; (c) a homoclinic loop, $\alpha=\ahom=0.2955$; and (d) the saddle's stable manifold, $\alpha=0.45$. Violet: basin of the origin; green: safe set with inactive filter ; beige: safe set with the filter active. Dark blue: $h=0$; dark blue dashed: switching line $\Sigma$. Blue dots and curves show initial conditions and trajectories. In (c), white dots mark the crossings $Q_k,Q_\alpha$ of the homoclinic loop with $\Sigma$.}
    \caption{Qualitative parameter map of the underdamped case, $\zeta=0.1$ (a), and phase portraits illustrating different regions of attraction at $k=-0.5$ (b)--(d), $d=1$. (b) $\alpha=0.20$: unstable periodic orbit. (c) $\alpha=\ahom=0.2955$: homoclinic loop. (d) $\alpha=0.45$: stable manifold of the saddle.}
    \label{fig:cell}
\end{figure*}

\subsection{Global stability for \texorpdfstring{$k\ge0$}{k >= 0}}\label{sec:ges}

In this section we study the case where $k \geq 0$.
First, note that both modes are Hurwitz when $k>0$. In general dimension, the sufficient condition of~\cite[Thm.~2]{MestresMousaviAmes2026} additionally requires a common quadratic Lyapunov function. 
The following result shows that in the plane, that extra assumption is unnecessary, and asymptotic recovery extends to $k=0$.

\begin{theorem}\longthmtitle{Global stability for $k\ge0$}\label{thm:ges}
Suppose that $\zeta>0$, $d>0$, and $\alpha>0$. If $k>0$, the origin of~\eqref{eq:nfcl} is globally exponentially stable (GES). If $k=0$, it is globally asymptotically stable (GAS), but not GES.
\end{theorem}
\begin{proof}
Let $k>0$ and set $q=\xi_1$, $v=\xi_2$. The filter raises the nominal acceleration only when required by $\dot h+\alpha h\ge0$, giving
\begin{equation}\label{eq:acceleration}
    \begin{aligned}
    \dot q&=v,\qquad \dot v=f(q,v),\\
    f(q,v)&=\max\{-q-2\zeta v,\,-k\alpha q-(k+\alpha)v-\alpha d\}.
    \end{aligned}
\end{equation}
Define the position-dependent term and its associated energy by
\[
    \begin{gathered}
    g(q)=-f(q,0)=\min\{q,k\alpha q+\alpha d\},\\
    E=\tfrac12v^2+\int_0^q g(s)\,ds.
    \end{gathered}
\]
Then $qg(q)\ge m q^2$, $m=\min\{1,k\alpha\}>0$, and $E$ has quadratic upper and lower bounds. The two velocity slopes give $c=\min\{2\zeta,k+\alpha\}>0$ and $C=\max\{2\zeta,k+\alpha\}$, so
\[
    \begin{aligned}
    \dot E&=vf(q,v)+g(q)v\le-cv^2,\\
    |f(q,v)-f(q,0)|&\le C|v|.
    \end{aligned}
\]
To obtain decay also in $q$, take $V=E+\varepsilon qv$~\cite{Koditschek1988}. Young's inequality yields
\[
    \dot V\le-\tfrac{\varepsilon m}{2}q^2
    -\bigl[c-\varepsilon(1+C^2/(2m))\bigr]v^2.
\]
Now, by taking  $\varepsilon < \frac{c}{1 + \frac{C^2}{2m} }$, we ensure $\dot{V} < 0$ whenever $(q,v)\neq(0,0)$,
proving GES.
For $k=0$, the same energy with $g(q)=\min\{q,\alpha d\}$ is radially unbounded, and LaSalle's principle gives global asymptotic stability. Convergence is locally exponential, but not globally exponential: far from the origin, the active dynamics admit arbitrarily long constant-speed returns with $v=-d$.
\end{proof}
%Theorem~\ref{thm:ges} shows that for $k>0$ the filter never destroys global recovery, whatever the gain $\alpha$. It assumes that $h$ has relative degree one. The next remark treats constraints of relative degree two, which correspond to $k\to\pm\infty$.

The following remark extends Theorem~\ref{thm:ges} to constraints of relative degree two.

\begin{remark}\longthmtitle{Relative degree two}\label{rem:reldeg2}
If $\bc^\top\bB=0$ but $\bc^\top \bA \bB \neq 0$, the high-order construction~\cite{XiaoBelta2022} uses the linear extended class-$\Kc$ function $s\mapsto\alpha_e s$, with slope $\alpha_e>0$, to define
\[
    h_e=\dot h+\alpha_e h
       =\bc^\top(\bA+\alpha_e\bI)\bx+\alpha_e d.
\]
% For a controllable planar pair, $\bc^\top\bA\bB\neq0$, and $h_e$ has relative degree one and zero $z_e=-\alpha_e<0$.
Enforcing $\dot h_e+\alpha h_e\ge0$ 
guarantees safety in the set $\setdef{x\in\real^2}{h(\bx) \geq 0, h_e(\bx) \geq 0}$.
Moreover, the matrix $\tilde{\bA}$ has eigenvalues $-\alpha, -\alpha_e$ and is therefore Hurwitz, which ensures that the origin is GES for every $\alpha,\alpha_e > 0$.
\demo
\end{remark}

\subsection{Region of attraction for \texorpdfstring{$k<0$}{k < 0}}\label{sec:roa}

The preceding subsection established global stability for $k\ge0$. For $k<0$, the origin is still locally exponentially stable, but a second equilibrium, the saddle $\xis=(-d/k,0)^\top$, lies on $h=0$ inside the active zone. Its stable manifold $\Ws(\xis)$ consists of initial conditions converging to the saddle rather than the origin.

Figure~\ref{fig:cell} shows how this additional equilibrium changes global dynamics. Panel~(a) locates three choices of gain at the same $k<0$, when the system is underdamped $0<\zeta<1$ and there is a stable focus. Panels~(b)--(d) show the corresponding trajectories. At the lowest gain, an unstable periodic orbit encloses the origin, meaning that trajectories inside return to the origin, whereas nearby trajectories outside move away from the orbit and escape. At the transition gain, this boundary reaches the saddle and becomes a homoclinic loop, a trajectory that leaves and returns to the same equilibrium. At higher gains, the closed boundary disappears, but convergence to the origin is still limited by the saddle's stable manifold. The violet area is the region of attraction in each panel. Its overlap with the safe set gives the initial conditions from which both safety and recovery are achieved.

For an underdamped nominal system, there is a transition at a specific gain at which a trajectory leaving the saddle returns to that same saddle. This trajectory forms the homoclinic loop, which defines the boundary between the periodic-orbit and stable-manifold cases. It leaves along $h=0$, crosses the inactive zone, and returns in the direction $\bv_\alpha$. Let $Q_k$ and $Q_\alpha$ be the intersections of $\Sigma$ with $h=0$ and with the line $\xis+\operatorname{span}\{\bv_\alpha\}$ respectively. As illustrated in Fig.~\ref{fig:cell}(c), a loop requires the nominal flow from $Q_k$ to reach $Q_\alpha$. With $D(s)=s^2-2\zeta s+1$, these points are
\begin{equation}\label{eq:crossings}
    Q_k=\frac{d}{D(k)}\begin{bmatrix}2\zeta-k\\-1\end{bmatrix},\quad
    Q_\alpha=\frac{\alpha d}{kD(\alpha)}\begin{bmatrix}2\zeta-\alpha\\-1\end{bmatrix}.
\end{equation}
The homoclinic orbit is defined when the nominal dynamics carry $Q_k$ to $Q_\alpha$, where the returning branch leads back to the saddle:
\[
    e^{\Az T}Q_k=Q_\alpha,
\]
with $T>0$ the time spent in the inactive zone. Substituting~\eqref{eq:crossings} and eliminating $T$ gives the scalar condition
\begin{equation}\label{eq:hom}
    \underbrace{\frac{\alpha^2D(k)}{k^2D(\alpha)}}_{L(\alpha)}
    =\underbrace{
    e^{-\frac{2\zeta}{\omega_{\rm d}}
    (2\pi-\theta(\alpha)+\theta(k))}
    }_{R(\alpha)},
\end{equation}
where $\omega_{\rm d}=\sqrt{1-\zeta^2}$ is the normalized damped natural angular frequency and $\theta(s)=\arctan((1-\zeta s)/(\omega_{\rm d}s))\in(-\pi/2,\pi/2)$.
At the homoclinic connection, $T=(2\pi-\theta(\alpha)+\theta(k))/\omega_{\rm d}$. % If $L<R$, the trajectory returns to $\Sigma$ outside $Q_\alpha$, and if $L>R$, it returns inside.
We denote the solution of~\eqref{eq:hom} by $\ahom$.
We note that equation~\eqref{eq:hom} (and hence $\ahom$) is independent of $d$.
However, $d$ scales the region of attraction of the origin.

\begin{theorem}\longthmtitle{Region of attraction for $k<0$}\label{thm:saddle}
Let $k<0$, $\zeta>0$, $d>0$, and $\alpha>0$. If $0<\zeta<1$, equation~\eqref{eq:hom} defines a unique homoclinic transition gain $\ahom\in(0,-k)$, with the following properties.
\begin{enumerate}
\item[(a)] If $0<\alpha<\ahom$, there is exactly one periodic orbit $\gu$, which is unstable. It surrounds the origin but not the saddle,
and $\Bc(\bzero)$ is its interior.
\item[(b)] If $\alpha=\ahom$, there is a homoclinic loop and no periodic orbit. The interior of the loop is $\Bc(\bzero)$.
\item[(c)] If $\alpha>\ahom$, there is no periodic orbit or homoclinic loop. The region of attraction is the component of $\real^2\setminus\Ws(\xis)$ containing the origin, and $\partial\Bc(\bzero)=\Ws(\xis)$.
\end{enumerate}
If $\zeta\ge1$, there is no periodic orbit or homoclinic loop, and $\Bc(\bzero)$ is as in (c). For $0<\zeta<1$, the safe recoverable set $\Bc(\bzero)\cap\Cc$ is bounded. For $\zeta\ge1$, it is unbounded whenever $\alpha\ge-\lambda_1$, where $\lambda_1=-\zeta+\sqrt{\zeta^2-1}$ is the slowest nominal pole. In every case, a trajectory outside $\overline{\Bc(\bzero)}\cup\Ws(\xis)$ diverges with $\lim\limits_{\tau\to\infty} h(\bxi(\tau))=0$, and it remains safe if it starts safe.
\end{theorem}
\begin{proof}
We first find the periodic orbits and homoclinic loops, then the region of attraction, and finally whether its safe part is bounded, using known results on planar piecewise linear systems.

\emph{Underdamped case.} Let $0<\zeta<1$. By~\eqref{eq:acceleration}, the damping is $2\zeta>0$ in the inactive zone and $k+\alpha$ in the active zone. For $\alpha\ge-k$ both are nonnegative, and Bendixson's criterion~\cite[Prop.~2.4]{PonceRosVela2022} excludes periodic orbits and homoclinic loops. For $0<\alpha<-k$, the results of~\cite[Prop.~20, Prop.~26]{Freire1998} and~\cite[Thm.~7.9, Prop.~7.7]{PonceRosVela2022}, whose condition for a homoclinic loop is~\eqref{eq:hom} in our notation, give a homoclinic loop if $L=R$, exactly one periodic orbit, which is unstable, if $L<R$, and neither if $L>R$. Since $L/R$ increases with $\alpha$, as $\frac{d}{d\alpha}\log(L/R)=2/(\alpha D(\alpha))>0$, from $0$ as $\alpha\to0^+$ to above $1$ at $\alpha=-k$, the equality $L=R$ holds at exactly one gain $\ahom\in(0,-k)$, with $L<R$ below it and $L>R$ above it.

\emph{Critically damped and overdamped case.} For $\zeta\ge1$, the nominal dynamics form a stable node. A trajectory crossing from the active into the inactive zone cannot cross back, since the Poincar\'e map from $\Sigma$ through the inactive zone back to $\Sigma$ is trivial~\cite[Prop.~4.5.2(b.1)]{LlibreTeruel2014}. It therefore converges to the origin. Hence there is no periodic orbit or homoclinic loop.

\emph{Region of attraction.} In case (a), trajectories inside $\gu$ stay inside it. They are therefore bounded, and since $\gu$ repels, they converge to the origin by Poincar\'e--Bendixson. Trajectories outside $\gu$ that do not lie on $\Ws(\xis)$ have no equilibrium or periodic orbit to converge to, so they are unbounded. They therefore enter the active zone beyond $Q_\alpha$, where their norm diverges while $h$ tends to zero~\cite[Prop.~4.4.15]{LlibreTeruel2014}. In case (b), the same argument applies to the homoclinic loop, whose inner side repels~\cite[Prop.~7.7]{PonceRosVela2022}. In case (c) and for $\zeta\ge1$, it applies to $\Ws(\xis)$: trajectories on the origin's side never reach the active zone beyond $Q_\alpha$, which lies on the other side, so they stay bounded~\cite[Prop.~4.4.15]{LlibreTeruel2014} and converge to the origin, while trajectories on the other side escape.

\emph{Bounded or unbounded safe recoverable set.} Let $\zeta\ge1$ and $\alpha\ge-\lambda_1$. Along the eigenvector of $\lambda_1$ the nominal motion decays at the rate $-\lambda_1\le\alpha$, so it satisfies the CBF condition~\eqref{eq:cbf} without the filter, and every safe state on this half-line converges to the origin. Hence $\Bc(\bzero)\cap\Cc$ is unbounded. Otherwise, either $0<\zeta<1$ and the nominal motion turns around the origin, or $\alpha<-\lambda_1$ and it approaches the slow eigenvector, along which the CBF condition fails far from the origin. In both cases a large safe state reaches the active zone far from the origin, beyond $Q_\alpha$, and escapes. Hence $\Bc(\bzero)\cap\Cc$ is bounded.
\end{proof}
\begin{figure}[t]
    \centering
    \includegraphics[width=\linewidth]{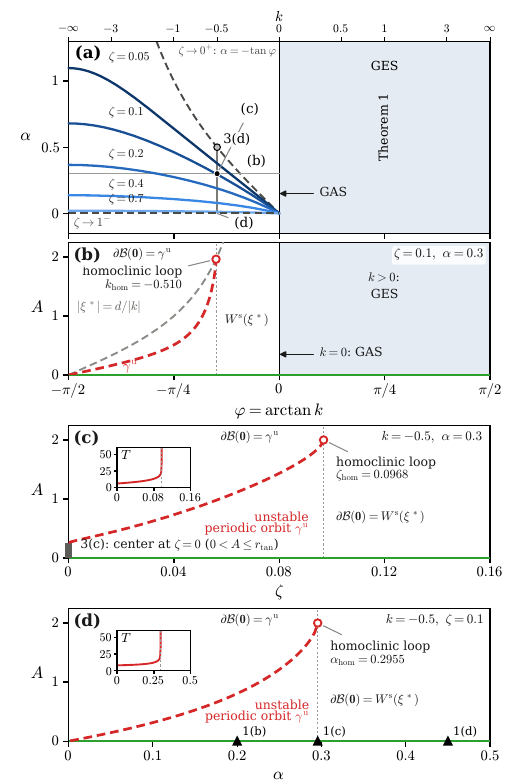}
%     \caption{Dependence of the basin boundary on the parameters with offset $d=1$. (a) The map in Fig.~\ref{fig:cell}(a) extended to several damping ratio $\zeta$ values. Dashed lines give the limits as $\zeta\to0^+$ and $\zeta\to1^-$. The horizontal line, the dot, and the vertical segment specify sweeps (b), (c), and (d). (b) Unstable periodic orbit amplitude versus $k$, $(\zeta,\alpha)=(0.1,0.3)$; grey dashed: saddle distance $|\xis|=d/|k|$. (c) Amplitude versus damping, $(k,\alpha)=(-0.5,0.3)$. The grey bar at $\zeta=0$ represents the center's closed trajectories in Fig.~\ref{fig:limit}(b). (d) Amplitude versus gain, $(k,\zeta)=(-0.5,0.1)$. Triangles mark the parameters corresponding to Fig.~\ref{fig:cell}(b)--(d). Insets in (c) and (d) show the period and green denotes the origin. Each branch ends at a homoclinic loop with $A=|\xis|$ and divergent period $T$.}
    \caption{Qualitative parameter map for several damping ratios (a) and bifurcation diagrams of the unstable periodic orbit $\gu$ with respect to $k$, $\zeta$, and $\alpha$ (b)--(d), $d=1$. The diagrams pass through $(k,\alpha,\zeta)=(-0.5,0.3,0.1)$, marked in (a), and each ends at a homoclinic loop. $A$ is the orbit amplitude, $T$ its period (insets), and the grey dashed curve in (b) is the saddle distance $|\xis|=d/|k|$.}
    \label{fig:bif}
\end{figure}

% Figure~\ref{fig:bif}(a) extends the parameter map of Fig.~\ref{fig:cell}(a) to several values of the damping ratio. 
Figure~\ref{fig:bif} shows how the region of attraction depends on the parameters. Its panel~(a) extends the parameter map of Fig.~\ref{fig:cell}(a) to several values of the damping ratio. 
Each curve gives the gain $\ahom(k,\zeta)$ resulting in homoclinic orbits. For a fixed $k<0$     parameter, increasing the gain crosses from an unstable periodic orbit to a homoclinic loop and then to a region of attraction bounded by the saddle's stable manifold. Increasing damping moves this transition to smaller gains. Thus, the same gain can give different regions of attraction for different nominal systems.

Panels (b)--(d) are bifurcation diagrams of the unstable periodic orbit $\gu$, while changing one parameter at a time. The vertical coordinate $A=\max_{\bxi\in\gu}\|\bxi\|_2$ gives the amplitude of the unstable periodic orbit bounding the region of attraction. In panel~(b), $\zeta=0.1$ and $\alpha=0.3$ are fixed while $k$ varies. The amplitude stays below the saddle distance $|\xis|=d/|k|$ and grows with $k$ until the orbit reaches the saddle at $k_{\rm hom}=-0.510$. For $k>0$ the origin is GES by Theorem~\ref{thm:ges}. In panel~(c), $k=-0.5$ and $\alpha=0.3$ are fixed while the damping varies. The orbit grows until it reaches the saddle at $\zeta_{\rm hom}=0.0968$. In panel~(d), the damping is fixed at $\zeta=0.1$ while the gain varies. Here, the transition occurs at $\ahom=0.2955$. The marked gains correspond to the three phase portraits of Fig.~\ref{fig:cell}(b)--(d). In all three diagrams the orbit ends in the homoclinic loop with finite amplitude, while its period diverges (insets), because the trajectory spends increasingly long near the saddle. Beyond the transition, $\Ws(\xis)$ bounds the region of attraction, and the origin (green) stays locally exponentially stable throughout.

\begin{figure}[t]
    \centering
    \includegraphics[width=0.95\linewidth]{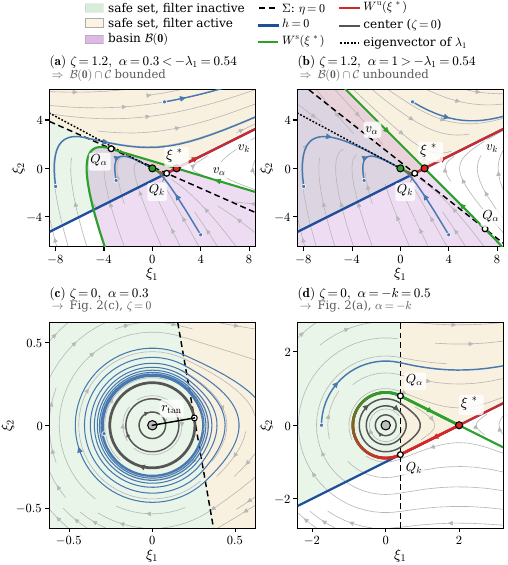}
    \caption{Phase portraits for an overdamped (a), (b) and an undamped (c), (d) nominal loop, $k=-0.5$, $d=1$. (a), (b) $\zeta=1.2$: the safe recoverable set is bounded for $\alpha=0.3$ and unbounded for $\alpha=1$. (c) $\zeta=0$, $\alpha=0.3$: circles around the center up to the one tangent to $\Sigma$, larger trajectories spiral outward. (d) $\zeta=0$, $\alpha=-k$: a center enclosed by a homoclinic loop of the saddle.}
    \label{fig:limit}
\end{figure}

The dashed limits in Fig.~\ref{fig:bif}(a) bound the underdamped range $0<\zeta<1$ of Theorem~\ref{thm:saddle}. As $\zeta\to1^-$, $\ahom\to0$, so the unstable periodic orbit exists only for smaller and smaller gains. For critically damped ($\zeta=1$) and overdamped ($\zeta>1$) nominal systems the poles are real, and there is no periodic orbit or homoclinic loop for any $\alpha>0$. Recovery is then separated from escape by $\Ws(\xis)$ alone. If, in addition, $\alpha\ge-\lambda_1$, the nominal motion along the slow eigenvector satisfies the CBF condition by itself, and the safe part of the region of attraction is unbounded, as in Fig.~\ref{fig:limit}(b). The filtered controller thus recovers some safe initial states arbitrarily far from the origin, but not every safe initial state. For smaller gains, as in Fig.~\ref{fig:limit}(a), and for every underdamped nominal loop, the safe part of the region of attraction is bounded (Theorem~\ref{thm:saddle}).

In the undamped case ($\zeta=0$) the origin is a center, and the nominal trajectories are circles around it. The circles that stay in the inactive zone never activate the filter, so they are unchanged up to the one that touches $\Sigma$, at radius
\begin{equation}\label{eq:grazing-radius}
    r_{\rm tan}=\frac{\alpha d}{\sqrt{(\alpha k-1)^2+(k+\alpha)^2}}.
\end{equation}
A trajectory starting just outside this circle enters the active zone, where the damping $k+\alpha$ in~\eqref{eq:acceleration} is negative for $\alpha<-k$. It therefore spirals outward, as illustrated by the trajectory outside the family of closed orbits in Fig.~\ref{fig:limit}(c). With small positive damping, the origin becomes attractive and an unstable periodic orbit appears near the circle of radius $r_{\rm tan}$~\cite[Thm.~3.28]{PonceRosVela2022}. This explains the nonzero starting amplitude $r_{\rm tan}=0.2570$ in Fig.~\ref{fig:bif}(c) (grey bar).

Finally, the limit $\ahom\to-k$ as $\zeta\to0^+$ leads to Fig.~\ref{fig:limit}(d). At exactly $\zeta=0$, $\alpha=-k$, both acceleration expressions in~\eqref{eq:acceleration} are independent of velocity, so energy is conserved. A homoclinic loop of the saddle then encloses the center, and the trajectories inside it keep circling the center instead of converging to the origin. This differs from Fig.~\ref{fig:cell}(c), where positive damping makes the interior of the loop the region of attraction.

\section{Example: vehicle sideslip protection}\label{sec:vehicle}

\begin{figure}[t]
    \centering
    \includegraphics[width=0.95\linewidth]{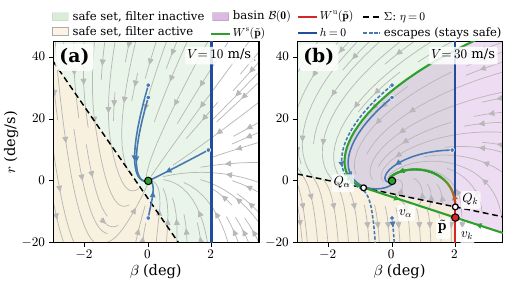}
    \caption{Phase portraits of the vehicle sideslip protection, $\beta_{\max}=2^\circ$, $\alpha=2~{\rm s^{-1}}$. (a) $V=10~{\rm m/s}<V_z$. (b) $V=30~{\rm m/s}>V_z$. Dashed trajectories escape while remaining safe.}
    \label{fig:vehicle}
\end{figure}

\begin{figure*}[t]
    \centering
    \includegraphics[width=\textwidth]{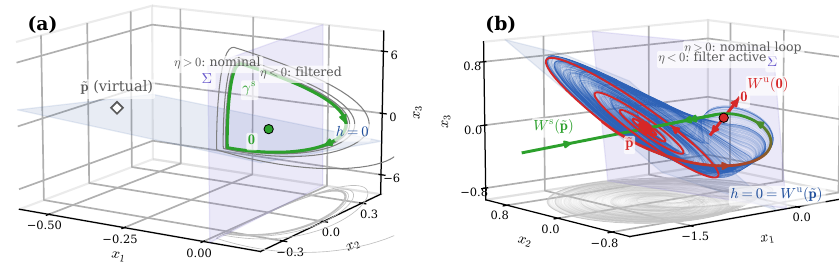}
    \caption{Phase portraits of two three-state examples of the form~\eqref{eq:companion}, $d=1$, with $\Sigma$ violet and $\{h=0\}$ blue. (a) $(a_0,a_1,a_2)=(7.5\cdot10^5,\,2.75\cdot10^4,\,300)$, $(c_0,c_1)=(2,3)$, $\alpha=5$: a solution (grey) converging to the stable limit cycle $\gamma^{\mathrm{s}}$ (green). (b) $(a_0,a_1,a_2)=(-0.55,\,0.615,\,0.15)$, $(c_0,c_1)=(0.9025,\,-0.1995)$, $\alpha=0.45$: the safe chaotic attractor (blue) near the homoclinic loop of the saddle-focus $\tp$ at $\ahom=0.4499$, drawn from red to green as it leaves $\tp$ along $\Wu(\tp)$ and returns along $\Ws(\tp)$.}
    \label{fig:3d}
\end{figure*}

We illustrate Theorems~\ref{thm:ges} and~\ref{thm:saddle} on the constant-speed bicycle model of a steer-by-wire vehicle~\cite{BobierTiu2019},
\begin{equation}\label{eq:vehicle}
\begin{aligned}
    \dot{\bx}&=\bA(V)\bx+\bB(V)\delta_f,\qquad \bB(V)=\begin{bmatrix}\frac{C_f}{mV}\\[2pt] \frac{aC_f}{I_z}\end{bmatrix},\\
    \bA(V)&=\begin{bmatrix}-\frac{C_f+C_r}{mV} & \frac{bC_r-aC_f}{mV^2}-1\\[2pt] \frac{bC_r-aC_f}{I_z} & -\frac{a^2C_f+b^2C_r}{I_zV}\end{bmatrix},
\end{aligned}
\end{equation}
where $\bx=(\beta,r)^\top$ collects the sideslip angle $\beta$ and the yaw rate $r$, and $\delta_f$ is the front steering angle. We use the parameters of the P1 test vehicle~\cite[Table~1]{BobierTiu2019}: $m=1725~{\rm kg}$, $I_z=1300~{\rm kg\,m^2}$, $a=1.35~{\rm m}$, $b=1.15~{\rm m}$, $C_f=75~{\rm kN/rad}$, $C_r=135~{\rm kN/rad}$. The open loop is Hurwitz at every speed, so we take $\bK=0$ and the filter only adds corrective steering. We protect one side of a sideslip envelope,
\begin{equation}\label{eq:vehicle-cbf}
    h(\bx)=\beta_{\max}-\beta\geq0,\qquad \beta_{\max}=2^\circ,
\end{equation}
so $\bc^\top=[-1\ \ 0]$ and $\bc^\top\bB=-C_f/(mV)\neq0$. The steering-to-sideslip zero
\begin{equation}\label{eq:vehicle-zero}
    z_\beta(V)=\frac{amV^2-bC_r(a+b)}{I_zV}
\end{equation}
changes sign at $V_z=\sqrt{bC_r(a+b)/(am)}=12.9~{\rm m/s}$. Below $V_z$ the constraint is minimum phase, $k>0$, and the filtered loop is GES by Theorem~\ref{thm:ges}. Above $V_z$ it is nonminimum phase, $k<0$, and the filter creates the saddle of Theorem~\ref{thm:saddle}, although the vehicle itself remains stable.

Figure~\ref{fig:vehicle} compares $V=10~{\rm m/s}$ and $V=30~{\rm m/s}$ at the same gain $\alpha=2~{\rm s^{-1}}$. At $10~{\rm m/s}$, $z_\beta=-11.9~{\rm s^{-1}}$, and every trajectory returns to straight driving with $\beta\le\beta_{\max}$. At $30~{\rm m/s}$ the nominal loop is a focus with $\omega_{\rm n}=8.54~{\rm s^{-1}}$ and $\zeta=0.711$, and $z_\beta=43.8~{\rm s^{-1}}$ gives $k=-5.13$. The saddle $\tp$ lies on the constraint at $(\beta,r)=(2^\circ,-11.9^\circ/{\rm s})$ with $\delta_f=-2.3^\circ$, a steady turn at $0.63\,g$. By~\eqref{eq:hom}, $\ahom=0.0182$, that is, $\omega_{\rm n}\ahom=0.155~{\rm s^{-1}}$ in physical time, well below $\alpha$, so $\Ws(\tp)$ bounds the region of attraction (Theorem~\ref{thm:saddle}(c)), and since the nominal loop is underdamped, its safe part is bounded. At $\beta=0$ the filtered vehicle recovers only from yaw rates between $-5^\circ/{\rm s}$ and $29^\circ/{\rm s}$. Outside the region of attraction the filter still keeps $\beta\le\beta_{\max}$, but it steers ever harder and the yaw rate diverges: safety is kept, recovery is lost.

\section{Beyond the plane}\label{sec:3d}

The planar analysis relies on the Poincar\'e--Bendixson theorem, which has no counterpart in higher dimensions. We therefore illustrate with two three-state examples what can change. The structure of the filtered loop is the same in any dimension: when the filter is active it enforces $\dot h=-\alpha h$, so the active dynamics have the eigenvalue $-\alpha$, and their remaining $n-1$ eigenvalues are the zeros of the constraint $(\bc^\top,\bA,\bB)$~\cite[Lemma~2]{MestresMousaviAmes2026}, see~\cite{MousaviMestresAmesMulti2026} for multiple constraints. Both examples use the controllable canonical form
\begin{equation}\label{eq:companion}
    \bA_0=\begin{bmatrix}0&1&0\\0&0&1\\-a_0&-a_1&-a_2\end{bmatrix},\quad
    \bB=\begin{bmatrix}0\\0\\1\end{bmatrix},\quad
    \bc=\begin{bmatrix}c_0\\c_1\\1\end{bmatrix},
\end{equation}
where the nominal poles are the roots of $s^3+a_2s^2+a_1s+a_0$ and the zeros of the constraint are the roots of $s^2+c_1s+c_0$.

The first example shows that Theorem~\ref{thm:ges} does not extend to three dimensions: two Hurwitz modes no longer guarantee global stability. In Fig.~\ref{fig:3d}(a) the nominal poles are $-50$, $-100$, $-150$ and the zeros are $-1$ and $-2$, so both modes are Hurwitz, and the origin is the only equilibrium. Nevertheless, all solutions we computed outside a narrow region around the origin converge to a stable limit cycle $\gamma^{\mathrm{s}}$. Since $\bA_0\tA$ has negative real eigenvalues, the two modes have no common quadratic Lyapunov function, so the sufficient condition of~\cite[Thm.~2]{MestresMousaviAmes2026} fails, and in $\real^3$ it cannot be dropped~\cite{CarmonaFreirePonceTorres2006}.

The second example shows that the filter can even produce chaos, here for an unstable nominal loop. In Fig.~\ref{fig:3d}(b) the nominal poles are $-0.35\pm0.94i$ and $0.55$, and the zeros of the constraint form an unstable complex pair $\lambda_z=0.10\pm0.94i$. The equilibrium $\tp$ created by the filter then has the eigenvalues $-\alpha$ and $\lambda_z$: it attracts along one direction and repels in a spiral within the plane $\{h=0\}$. At $\alpha=\ahom=0.4499$, a trajectory that spirals away from $\tp$ within $\{h=0\}$ leaves the region where the filter is active, travels through the region where it is inactive, and comes back to $\tp$ along the attracting direction. This trajectory is a homoclinic loop, drawn from red to green in Fig.~\ref{fig:3d}(b). By Shilnikov's theorem~\cite[Ch.~13]{ShilnikovBook2001},~\cite[Ch.~6]{Kuznetsov2004}, which also holds for continuous piecewise-linear systems~\cite{LlibrePonceTeruel2007}, such a loop generates infinitely many periodic orbits and chaotic dynamics when the real eigenvalue dominates the spiral, $\alpha>\operatorname{Re}\lambda_z$, here $0.45>0.10$. The attractor at $\alpha=0.45$ is chaotic, with largest Lyapunov exponent $0.085$, and safe: it grazes the constraint but never crosses it.

This mechanism is not available under the assumptions of Section~\ref{sec:model}. For a Hurwitz $\bA_0$ and $d>0$, $\tp$ is an equilibrium only if $\tA$ has an odd number of positive real eigenvalues~\cite[Prop.~2]{MestresMousaviAmes2026}, so the zeros are real, $\tp$ does not spiral, and the homoclinic loop behind the chaos of Fig.~\ref{fig:3d}(b) cannot form. A stable nominal loop in $\real^3$ can still oscillate, as in Fig.~\ref{fig:3d}(a), or have unbounded solutions~\cite[Fig.~2]{MestresMousaviAmes2026}. The full three-dimensional classification is left to the extended version of this work.

\section{Conclusion}\label{sec:conclusion}

We showed that a CBF safety filter can change the global behavior of an otherwise stable planar loop: it never violates the constraint, but it can create an additional equilibrium, oscillations, and diverging trajectories. Whether this happens is decided by a single property of the constraint, the sign of its zero. For a minimum-phase constraint the origin is recovered from every initial condition and for every gain, while for a nonminimum-phase constraint recovery holds only on a region whose boundary we characterized as an unstable periodic orbit, a homoclinic loop, or the saddle's stable manifold, depending on the gain and the nominal damping. Outside this region the filter still keeps the system safe, but the state grows along the constraint boundary. Recovery from arbitrarily large safe states is possible only if the nominal loop does not oscillate and the gain is at least the decay rate of its slow mode.
In practice, this suggests checking the phase of the constraint before relying on a safety filter for recovery. In vehicle sideslip protection, for instance, the constraint becomes nonminimum phase above a critical speed. Beyond the plane the behavior is richer: two Hurwitz modes can sustain a stable oscillation, and with an unstable plant the filter can produce a safe chaotic attractor. Extensions to higher dimensions, unstable plants, constraints that exclude the origin, and controller synthesis that shapes the region of attraction are left to future work.

\bibliographystyle{IEEEtran}
\bibliography{refs}

@article{MousaviMestresAmesMulti2026,
  author  = {Shima Sadat Mousavi and Pol Mestres and Aaron D. Ames},
  title   = {Stability Analysis in Multi-Constraint Safety Filters
             for Linear Systems},
  journal = {arXiv preprint arXiv:2606.15942},
  year    = {2026}
}

@article{BobierTiu2019,
  author  = {Carrie G. Bobier-Tiu and Craig E. Beal and
             John C. Kegelman and Rami Y. Hindiyeh and
             J. Christian Gerdes},
  title   = {Vehicle Control Synthesis Using Phase Portraits
             of Planar Dynamics},
  journal = {Vehicle System Dynamics},
  volume  = {57},
  number  = {9},
  pages   = {1318--1337},
  year    = {2019},
  doi     = {10.1080/00423114.2018.1502456}
}

@article{AmesTAC2017,
  author  = {Ames, Aaron D. and Xu, Xiangru and Grizzle, Jessy W. and Tabuada, Paulo},
  title   = {Control barrier function based quadratic programs for safety critical systems},
  journal = {IEEE Transactions on Automatic Control},
  volume  = {62},
  number  = {8},
  pages   = {3861--3876},
  year    = {2017},
  doi     = {10.1109/TAC.2016.2638961}
}

@inproceedings{AmesECC2019,
  author    = {Ames, Aaron D. and Coogan, Samuel and Egerstedt, Magnus and Notomista, Gennaro and Sreenath, Koushil and Tabuada, Paulo},
  title     = {Control barrier functions: Theory and applications},
  booktitle = {Proc. 18th European Control Conference (ECC)},
  address   = {Naples, Italy},
  pages     = {3420--3431},
  year      = {2019},
  doi       = {10.23919/ECC.2019.8796030}
}

@article{XiaoBelta2022,
  author  = {Xiao, Wei and Belta, Calin},
  title   = {High-order control barrier functions},
  journal = {IEEE Transactions on Automatic Control},
  volume  = {67},
  number  = {7},
  pages   = {3655--3662},
  year    = {2022},
  doi     = {10.1109/TAC.2021.3105491}
}

@article{ReisAguiarTabuada2021,
  author  = {Reis, Matheus F. and Aguiar, A. Pedro and Tabuada, Paulo},
  title   = {Control barrier function-based quadratic programs introduce undesirable asymptotically stable equilibria},
  journal = {IEEE Control Systems Letters},
  volume  = {5},
  number  = {2},
  pages   = {731--736},
  year    = {2021},
  doi     = {10.1109/LCSYS.2020.3004797}
}

@article{TanDimarogonas2024,
  author  = {Tan, Xiao and Dimarogonas, Dimos V.},
  title   = {On the undesired equilibria induced by control barrier function based quadratic programs},
  journal = {Automatica},
  volume  = {159},
  pages   = {111359},
  year    = {2024},
  doi     = {10.1016/j.automatica.2023.111359}
}

@inproceedings{ChenMestresCDC2024,
  author    = {Chen, Yiting and Mestres, Pol and Dall'Anese, Emiliano and Cort\'es, Jorge},
  title     = {Characterization of the dynamical properties of safety filters for linear planar systems},
  booktitle = {Proc. 63rd IEEE Conference on Decision and Control (CDC)},
  address   = {Milan, Italy},
  pages     = {2397--2402},
  year      = {2024},
  doi       = {10.1109/CDC56724.2024.10885923}
}

@article{MestresJNLS2026,
  author  = {Mestres, Pol and Chen, Yiting and Dall'Anese, Emiliano and Cort\'es, Jorge},
  title   = {Control barrier function-based safety filters: Characterization of undesired equilibria, unbounded trajectories, and limit cycles},
  journal = {Journal of Nonlinear Science},
  volume  = {36},
  number  = {2},
  pages   = {41},
  year    = {2026},
  doi     = {10.1007/s00332-026-10255-x}
}

@article{MestresMousaviAmes2026,
  author  = {Mestres, Pol and Mousavi, Shima Sadat and Ames, Aaron D.},
  title   = {Dynamical properties of safety filters for linear systems and affine control barrier functions},
  journal = {IEEE Control Systems Letters},
  volume  = {10},
  pages   = {979--984},
  year    = {2026},
  doi     = {10.1109/LCSYS.2026.3704420}
}

@misc{ChoiMinPhase2025,
  author        = {Choi, Jason J. and Tomlin, Claire J. and Sastry, Shankar and Sreenath, Koushil},
  title         = {When are safety filters safe? {O}n minimum phase conditions of control barrier functions},
  howpublished  = {arXiv:2508.07684},
  year          = {2025},
  note          = {Submitted to the IEEE for possible publication}
}

@inproceedings{MarcheseECC2025,
  author    = {Marchese, Nicola and Carrasco, Joaquin and Seiler, Peter and Zhang, Kaiqiang},
  title     = {High-order control barrier functions: Safety can lead to instability},
  booktitle = {Proc. European Control Conference (ECC)},
  address   = {Thessaloniki, Greece},
  pages     = {2148--2153},
  year      = {2025},
  doi       = {10.23919/ECC65951.2025.11186946}
}

@article{Freire1998,
  author  = {Freire, Emilio and Ponce, Enrique and Rodrigo, Francisco and Torres, Francisco},
  title   = {Bifurcation sets of continuous piecewise linear systems with two zones},
  journal = {International Journal of Bifurcation and Chaos},
  volume  = {8},
  number  = {11},
  pages   = {2073--2097},
  year    = {1998},
  doi     = {10.1142/S0218127498001728}
}

@book{PonceRosVela2022,
  author    = {Ponce, Enrique and Ros, Javier and Vela, Elisabet},
  title     = {Bifurcations in Continuous Piecewise Linear Differential Systems: Applications to Low-Dimensional Electronic Oscillators},
  series    = {RSME Springer Series},
  volume    = {7},
  publisher = {Springer},
  address   = {Cham},
  year      = {2022},
  doi       = {10.1007/978-3-031-21135-5}
}

@book{Kuznetsov2004,
  author    = {Kuznetsov, Yuri A.},
  title     = {Elements of Applied Bifurcation Theory},
  edition   = {3rd},
  series    = {Applied Mathematical Sciences},
  volume    = {112},
  publisher = {Springer},
  address   = {New York},
  year      = {2004},
  doi       = {10.1007/978-1-4757-3978-7}
}

@inproceedings{Koditschek1988,
  author    = {Koditschek, Daniel E.},
  title     = {Strict global {Lyapunov} functions for mechanical systems},
  booktitle = {Proc. American Control Conference},
  address   = {Atlanta, GA, USA},
  pages     = {1770--1775},
  year      = {1988},
  url       = {https://kodlab.seas.upenn.edu/uploads/Kod/Lyapunov_1988.pdf}
}

@article{CarmonaFreirePonceTorres2006,
  author  = {Carmona, Victoriano and Freire, Emilio and Ponce, Enrique and Torres, Francisco},
  title   = {The continuous matching of two stable linear systems can be unstable},
  journal = {Discrete and Continuous Dynamical Systems},
  volume  = {16},
  number  = {3},
  pages   = {689--703},
  year    = {2006},
  doi     = {10.3934/dcds.2006.16.689}
}

@book{ShilnikovBook2001,
  author    = {Shilnikov, Leonid P. and Shilnikov, Andrey L. and Turaev, Dmitry V. and Chua, Leon O.},
  title     = {Methods of Qualitative Theory in Nonlinear Dynamics, Part {II}},
  publisher = {World Scientific},
  address   = {Singapore},
  year      = {2001},
  series    = {World Scientific Series on Nonlinear Science, Series A},
  volume    = {5}
}

@article{LlibrePonceTeruel2007,
  author  = {Llibre, Jaume and Ponce, Enrique and Teruel, Antonio E.},
  title   = {Horseshoes near homoclinic orbits for piecewise linear differential systems in {$\mathbb{R}^3$}},
  journal = {Int. J. Bifurcation Chaos},
  volume  = {17},
  number  = {4},
  pages   = {1171--1184},
  year    = {2007}
}

@misc{LawanCarrascoSu2026,
  author        = {Lawan, Faisal and Carrasco, Joaquin and Su, Lanlan},
  title         = {Safe stabilising full-order affine control barrier functions for linear systems (extended)},
  howpublished  = {arXiv:2609.12990},
  year          = {2026}
}

@misc{MestresMousaviOngAmes2026,
  author        = {Mestres, Pol and Mousavi, Shima Sadat and Ong, Pio and Ames, Aaron D.},
  title         = {Safe stabilizing linear feedback: {N}ecessary and sufficient conditions, optimality, and margins},
  howpublished  = {arXiv:2607.21794},
  year          = {2026}
}

@book{Chen1999,
  author    = {Chen, Chi-Tsong},
  title     = {Linear System Theory and Design},
  edition   = {3rd},
  publisher = {Oxford University Press},
  address   = {New York, NY},
  year      = {1999}
}

@article{Kiss_CBFal,
author = {Kiss, Adam K. and Molnar, Tamas G. and Ames, Aaron D. and Orosz, Gabor},
title = {Control barrier functionals: Safety-critical control for time delay systems},
journal = {International Journal of Robust and Nonlinear Control},
volume = {33},
number = {12},
pages = {7282-7309},
doi = {https://doi.org/10.1002/rnc.6751},
year = {2023}
}

@book{LlibreTeruel2014,
  author    = {Llibre, Jaume and Teruel, Antonio E.},
  title     = {Introduction to the Qualitative Theory of Differential Systems: Planar, Symmetric and Continuous Piecewise Linear Systems},
  publisher = {Birkh\"auser},
  address   = {Basel},
  year      = {2014},
  doi       = {10.1007/978-3-0348-0657-2}
}

@article{Mousavi2026structure,
  title={Structure, feasibility, and explicit safety filters for linear systems},
  author={Mousavi, Shima Sadat and Cohen, Max H and Mestres, Pol and Ames, Aaron D},
  journal={arXiv preprint arXiv:2604.04235},
  year={2026}
}

\end{document}